# Sensory evaluation of commercial coffee brands in Colombia

## Edis Mauricio Sanmiguel Jaimes

Departamento de Administración de empresas,
Universidad Libre – Colombia – seccional Socorro,
Calle 21 No. 7-75 San Gil, Santander, Colombia
Email: ingedis@gmail.com

## Igor Barahona Torres*

Department of Statistics, Mathematics and Computer Science,
Cátedras CONACYT,
Autonomous University of Chapingo,
56230 Km. 37.5 highway México-Texcoco, Mexico
Email: igor0674@gmail.com
*Corresponding author

## Héctor Hugo Pérez-Villarreal

Department of Management and Marketing,
Popular Autonomous University of Puebla State,
17 Sur 901, Barrio de Santiago, Puebla, Mexico
Email: hectorhugo.perez@upaep.mx

**Abstract:** Colombian coffee farmers have traditionally focused their efforts on activities including seeding, planting and drying. Strategic issues to successfully compete in the industry, such as branding, marketing and consumer research, have been neglected. In this research, we apply a type of sensory analysis, based on several statistical techniques used to investigate the key features of ten different brands of Colombian coffee. A panel composed of 32 judges investigated nine different attributes related to flavour, fragrance, sweetness and acidity, among others. The last section presents the conclusions reached regarding customer preference and brands profiles.

## 1   Introduction

It is extremely difficult to provide a formal definition of the concept of quality, as it is "a complex and multidimensional concept that incorporates an infinity of attributes that configure and condition it, being particular for each of the product category or service." According to Calvo Fernández (1997), some features that are found within the concept of quality can distinguish a target quality (located on the side of supply), and a perceived quality (located on the side of the application). The first attempts to measure the target and verify the characteristics of the product in terms of the aspects that configure them, while the second refers to the consumers' appreciation of these characteristics, expressed by the difference between what is expected and what is actually obtained (Calvo Fernández, 1997). Literature shows various approaches to the concept of product quality; four approaches to the same concept can be considered:

1   the metaphysical philosophical approach

2   the production management approach

3   the economic approach

4   the focus on perceived quality or marketing performance and consumer behaviour (Garvin, 1984, cited by Steemkamp, 1989).

The food markets distinguished here offer products obtained through the use of technology and inputs following authorised processes and rules implemented to assure 'quality'. The growing demand for high quality products is due to the attributes that they possess to satisfy consumer perceptions on health, safety and other relevant features in their purchasing decisions (Lacaze et al., 2007); the natural environment, through factors such as soil, geography, topography, weather and crops; as well as historical and cultural specificities through the production and processing methods used to configure the characteristics and the quality of many agro-food products (Couillerot, 2000, cited by Granados and Alvarez, 2002).

Based on this general perspective of the concept of quality, this article will focus on the quality of Colombian coffees. The main objective of this research is to provide empirical evidence that will assist producers to identify the strengths and weaknesses on their products, as it simultaneously generates valuable information related to the most important coffee brands in Colombia, in order to assist customers to make better informed purchase decisions. To achieve this objective, we asked three basic questions:

1  Are the ten most important Colombian coffee brands perceived as equal or different?

2  How can a specific brand of Colombian coffee be characterised among a group?

3  Is there a consensus of a group of judges regarding with a given attribute?

The rest of this article is composed as follows: the next section offers a discussion of sensory evaluation methods applied to coffee; Section 3 provides the formal definitions of the descriptors used to evaluate the coffees in this research; Section 4 presents a description of the quantitative model; market segmentation and features related to the evaluating panel are discussed in Section 5; with the results are presented on Section 6. The last section is reserved for the conclusions.

## 2  The sensory evaluation of coffee

The US Institute of Food Technologists (IFT) defines sensory evaluation as "the scientific discipline used to evoke, measure, analyse and interpret the reactions to those characteristics of foods and other substances that are perceived by the senses of sight, smell, taste, touch and hearing." On the other hand, sensory analysis is a scientific discipline that is useful in the measurement, analysis and interpretation of the complex feelings experienced by people with regard to certain characteristics of foods (Calí, 2002, cited by Césari et al., 2013). Sensory evaluation uses one or more of the five senses to evaluate foods. The tasting panels, formed of a group of people, evaluate specific food samples under controlled conditions to evaluate them in different ways, depending on the specific sensory test realised. This is a unique type of test that can measure consumer acceptance and preference. To know the public opinion about a product, there is no substitute for the evaluation by individual consumers (Vaclavik, 1998).

Sensory evaluation allows the characterisation of a batch of coffee. The cup test assesses the characteristics of a certain coffee by tasters as they seek to recognise its strengths and weaknesses (Zapata Gómez and Sarache Castro, 2014). Coffee consumers prefer products of the highest quality, not only good flavour and taste, but also that its consumption does not affect consumer's health. The quality of a coffee beverage depends on many factors: genetic origin, latitude, altitude, local weather, farming, healthcare, agronomic practices, coffee culture, crop quality, type and control during the beneficiation, harvesting, storage, roasting and beverage preparation processes (Puerta Quintero, 1988). Tasting is the method used to determine the characteristics of aroma, taste and coffee health. This test, also referred to as the sensory quality evaluation of coffee and cup test, is used to identify any defects present in the coffee beverage, to measure the intensity of a sensory characteristic like acidity and sweetness, and likewise to describe the flavour, aroma and global quality of the product (Puerta Quintero, 2009). The sensory evaluation of food quality is similarly used as a form of evaluating and describing the product categories, as well as implementing sets of quantitative scales. For this reason it is also called the quantitative descriptive method. The scale proposed by Puerta Quintero (1997), entitled 'Scale for the assessment of the quality of the green arabica coffee beverage' represents an example of a quantitative scale applied on assessing beverages. One of the advantages of sensory analysis is its reproducibility, and its capacity to describe and characterise food and beverages. This is a method that allows an individual as well as integral evaluation of the sensory characteristics of the coffee and the relationships between them, in order to obtain the greatest amount of information possible from a single sample (ICONTEC, 2011).

## 3 Descriptors used in the sensory evaluation of coffee

A descriptor is a term that refers to a product characteristic to be evaluated. The properties of the descriptor must be such that they can be evaluated on an intensity scale (ICONTEC, 2011). The sensory qualities of coffee: aroma, acidity, bitterness, body, flavour, health and grain quality, are the most important aspects in the acceptance and definition of coffee quality (Puerta Quintero, 1997). According to INFOCAFES (2009), the descriptors that should be considered in a sensory evaluation of coffee, are briefly described below:

- *Fragrance*: the intensity produced by the volatile components of coffee that are perceived by smell without the addition of water (ICONTEC, 2011). This is the first indicator of the sample quality, however, it should not be qualified as a separate item regardless of the aroma. From this beginning, positive or negative attributes of the coffee analysed can be manifested to the tester (USAID, 2005).

- *Aroma*: corresponds to the olfactory perception of intensity and quality, due to the volatile compounds of coffee entrained by water steam at the time of making the beverage (FEDECAFE, 2006). A good coffee drinker, like a wine taster, will smell/test the aroma before wetting his lips with the coffee. What we taste is determined by what we smell (INFUSIONISTS, 2014). As for USAID (2005), the smell, this is the smell of coffee and gives us a general idea of the already ground sample once water is added.

- *Acidity*: sensation caused by the acidic substances of coffee that pass to the drink and are detected in the oral cavity. Depending on variety, profit system, growing region, degree of roast and raw materials, the higher the degree of roasting, the less the acidity (ICONTEC, 2011). Acidic intensity is a desirable feature, appreciated in coffee, as there is a positive relationship between the intensity of the acid impression and coffee quality (INFOCAFES, 2009).

- *Bitterness*: feeling produced by bitter coffee substances when they are extracted with hot water and passed to the beverage. Bitterness is perceived mostly at the back of the tongue. An unpleasant bitterness can be due to an excess coffee extraction. The higher the degree of roasting the more bitter taste (ICONTEC, 2011)

- *Body*: strength, character and heaviness of the beverage associated with the characteristics of the raw material and the water-coffee ratio employed in preparing the beverage (ICONTEC, 2011). Body is the thickness of the flavour, consistency or thickness of the liquid (USAID, 2005).

- *Sweetness*: primary olfactory and gustatory sensation perceived primarily in the fungiform papillae of the tip of the tongue, and associated with the presence of sweeteners. Soft, clean sensation, pleasant, in the coffee beverage.

- *Aftertaste*: Post-tasting flavours that remain in the mouth after tasting the drink. These may be pleasant or unpleasant depending on the initial features of the raw materials, processing and product storage conditions (ICONTEC, 2011). Residual taste/post taste flavour is the permanence of flavour on the palate after having expulsed the coffee from the mouth. This can be agreeable, leaving a sweet and refreshing taste; or disgusting, leaving a bitter or harsh taste (USAID, 2005)

- *Roasting-Agtron*: The degree of the coffee roast is determined qualitatively by colour. Coffee origin and properties may influence the shades obtained during the roasting process. In general terms, the clearer the colour the lower the roast, and the softer, more acidic and less bitter the flavour. The darker the coffee the higher the roast, with a stronger, less acidic and more bitter flavour (FEDECAFE, 2006). Some colour ranges have been defined for coffee roast and ground coffee to express the degree of roasting. For instance, the Specialty Coffee Association of America (SCAA) has developed a point system to grade the colour level for different types of roast (SCAA, 2009). The system consists of eight numbered disks that represent a certain colour. Ground coffee is pressed into a Petri dish and the colour compared against the disks. A number is then assigned to the degree of roast according to the Agtron Gourmet Scale which ranges from no. 95 (the lightest roast), and then lower at intervals of from 10 to 25 as the roast darkens (The Coffee Guide, 2014).

- *Global impression*: Note made by the judge at the time of judging the beverage. Summary and group judgement regarding the quality of the coffee sample analysed (ICONTEC, 2011)

## 4 Model description

The generalised linear model (GLM) approach, introduced by Nelder and Wedderburn (1972) is applied here to analyse our collected data. According to Calcagno and Mazancourt (2010), GLMs provide a flexible framework to describe how a dependent variable is explained in terms of a set of explanatory variables (sometimes called predictors). GLMs also offer an additional advantage as dependent variables can be either continuous or discrete. On the other hand, the explanatory variables can be either quantitative or categorical. According to Calcagno and Mazancourt (2010), categorical variables are referred to as 'covariates', while categorical descriptors are referred to as 'factors'. According to Nelder and Wedderburn (1972), GLMs receive the name linear as they are assumed to have a linear effect on the transformation of the dependent variable defined by the link function and random effect on the error part of the model. That is to say, the term 'linear model' encompasses both systematic and random components. It is composed of $x_i$ independent variables whose values are supposedly known, an equal number of coefficients or parameters (for example $\alpha_i$, $\beta_i$, $\gamma_i$, etc.) and the model error (usually represented as $\varepsilon_i$). The coefficients are likewise supposedly fixed and unknown, and therefore must be estimated. Breslow (1996) identifies four basic assumptions behind GLM's. First, all observations must be statistically independent; residuals must be randomly distributed; the variance structure correctly specified and a linear relationship between the independent variable and factor. Figure 1 shows how the assumptions were validated for the factor 'judge' in our model. Validations for all other factors of the model are omitted here due to space limitations, however all were similarly calculated.

**Figure 1**  Validating assumptions of the model for the factor judge (see online version for colours)

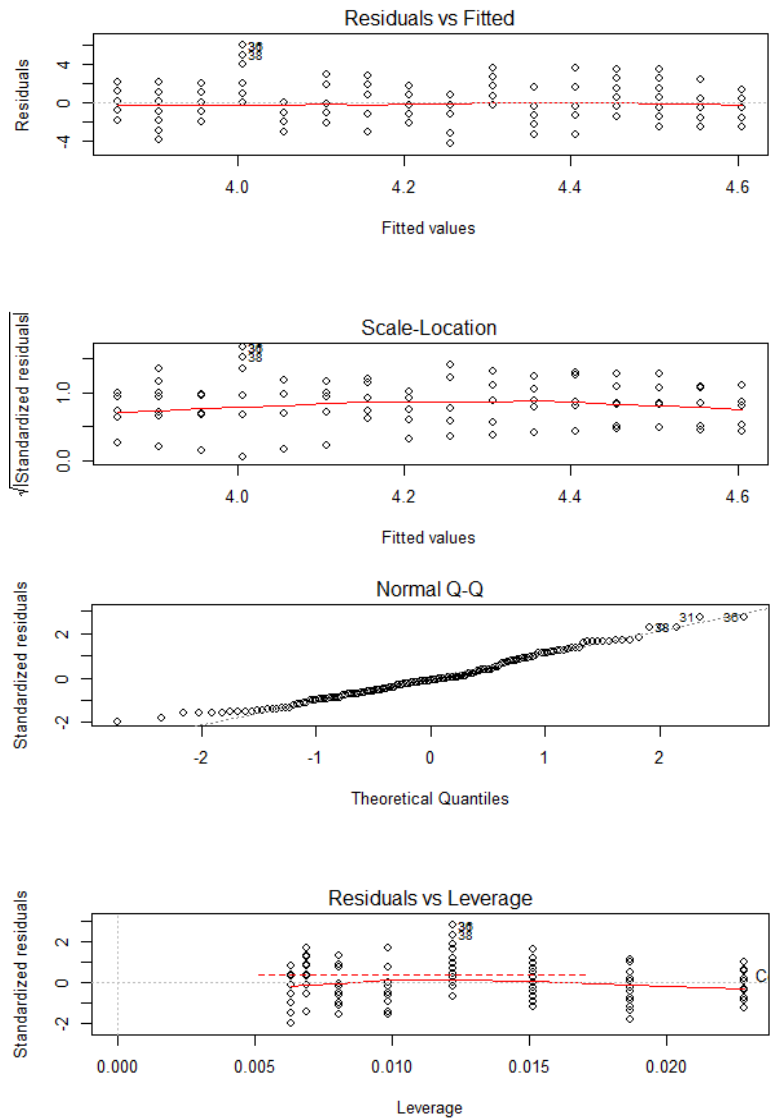

With regard to the structure of the model, three factors are analysed for the 320 observations that comprise a dataset. Namely, if 32 panellists participated in our study and each panellist evaluated ten different brands of coffee using nine different criteria, then our dataset is on the order of 320 × 9. Formulation (1) presents the linear model with the three factors of coffee, judge and session. As previously mentioned, the factor 'coffee' represents ten different brands of Colombian coffee, 'judge' refers to either graduate or undergraduate students and, the 'session' is the number of replicas carried out in the study. The model also considers three types of interactions between the main effects, which are coffee-judge, coffee-session and judge-session.

$$Y_{ijk} = \mu + \alpha_i + \beta_j + \gamma_k + \alpha\beta_{ij} + \alpha\gamma_{ik} + \beta\gamma_{jk} + \varepsilon_{ijk} \qquad (1)$$

$$L(\varepsilon_{ijk}) = N(0, \sigma) \quad \text{cov}(\varepsilon_{ijk}, \varepsilon_{i'j'k'}) = 0$$

$\mu$ mean effect

$\alpha_i$ coffee effect

$\beta_j$ judge effect

$\gamma_k$ session effect

$\alpha\beta_{ij}$ product-judge interaction

$\alpha\gamma_{ij}$ product-secion interaction

$\beta\gamma_{jk}$ judge-secion interaction

$\varepsilon_{ijk}$ model error.

With regard to the type of effect from each factor, coffee is defined as a fixed effect considering that there is a fixed number of brands participating on the study. The judge factor is considered as a random effect as the participating students were selected randomly. Finally, the session effect is a fixed effect given that only one replica was conducted. It is important to reiterate that based on the lack of replicas it was not possible to estimate the effect of the coffee-judge interaction.

## 5 Market segmentation and panel features

In the Colombian coffee industry like in the global industry, there are different manufacturing processes that can be classified into two major groups: roasted coffee (whole or ground) and soluble (instant) coffees. Natural roasted coffee is roasted without any additives by subjecting it to a stream of hot air. This causes the bean to grow considerably and to lose about 18% of its weight (Food Industry, 2014). Soluble coffee (lyophilised or solubilised) is obtained by dehydration or by drying the coffee. Ground and roasted coffees are used as reference here, as they correspond to the highest consumption segment in the country – about 86% of all coffee consumed in the entire country (pyme.com, 2013). Decaffeinated and flavoured coffees were discarded from our study due to their low consumption. Another aspect taken into account is the quality characteristics presented on the product label by the producer and for the end consumer. A segmentation is proposed in this respect considering the quality aspects of the raw material used to make the coffee. Market price also becomes an indicator of product quality, while another aspect corresponds to the geographical denomination of the coffee, i.e., the region of the country from where it comes. A total of six segments were identified: traditional, flavoured and decaffeinated, premium or specialty, with designation of origin, with social content and gourmet. This segmentation was developed after reviewing the shops, supermarkets and convenience stores located in the municipalities of Socorro and San Gil in the Department of Santander, Colombia.

The judges evaluated physical descriptors such as degree of roasting using the Agtron® colour scale (SCAA, 2009; Davids, 2010) in conjunction with the sensory descriptors discussed in the previous section. The techniques proposed by the national coffee research centre and other authors' leaders in the field (Puerta Quintero, 1998) were adapted in order to obtain valid and accurate results in the sensory analysis. Best practices documented by the SCAA (2009), the Colombian technical standard NTC 4883 Sensory Analysis and ICONTEC (2011) were also incorporated during the data collection process. Table 1 provides a summary of the variables incorporated in the study.

The panels were structured applying criteria that classified the participating students according to their studies and classifying them into two subgroups:

1  undergraduate

2  graduate students.

First a group composed of 16 undergraduate students evaluated the ten coffee brands using the descriptors presented in Table 1. A second evaluation was made by the graduate students, maintaining the same conditions and variables. On both cases, a ten-level quantitative scale was applied, on which 1 represented 'worst' and ten 'best'. Data were collected on sheets of paper and later captured in an Excel spreadsheet. The electronic spreadsheet was then exported to the open-access software SensoMiner® (2015), in order to run the calculations and the analysis (more details about the software can be found at http://sensominer.free.fr/).

**Table 1**    Input variables in the study for the characterisation of products

| Coffee descriptors | Number of panellists | Number of evaluated brands | Number of segments |
|---|---|---|---|
| Fragrance | 32 | 10 | 6 |
| Aroma | | | |
| Acidity | | | |
| Bitterness | | | |
| Body | | | |
| Sweetness | | | |
| Aftertaste | | | |
| Roasting-Agtron | | | |
| Global impression | | | |

Considering that roasted coffee, both whole and ground represents 86% of the country's trade (Portfolio, 2012), this was used as the baseline for the research project. The project was then conducted by classifying the six market segments with roasted and ground coffee, considering product quality, expressed as the raw material and the market price. The category of decaffeinated and flavoured coffee was discarded because the process requires altering the natural conditions of the roasted and ground coffee manufacturing process. As a result, we received the following proposed segmentation, which is also used as reference in this research. (Note that all the following prices are in Colombian pesos).

- *Traditional coffee*: this is a product made from a mixture of low-quality raw materials corresponding to the product segment, and that are the most popular in the market. The prices for these products range from $6,000 to $9,000 per pound.

- *Premium or specialty*: commercial products that use raw materials such as specialty or premium coffee. Prices range from $9,100 to $12,000 per pound.

- *Gourmet*: coffee with raw materials that meet special conditions such as aromatic fruity notes r herbal and that corresponds to high quality raw materials. Prices range from $12,100 to $16,000 per pound.

- *Designation of origin*: corresponds to products with an appellation of origin for a specific geographical area. In most cases, this is given by environmental or geographical conditions. Prices for this segment can range from $16,100 to $20,000 per pound.

- *Sustainable and organic (or social content coffees)*: corresponds to products that have been certified, such as for instance, 'bird-friendly', 'rainforest alliance', 'USDA organic' and 'fair trade', among others (more information related with coffee certifications can be reached on http://www.ico.org/sustaininit.asp). The price range for this product is from 16,000 to 21,000 per pound).

- *Flavoured and decaffeinated coffees*: Flavoured coffees have added ingredients that complement the coffee taste. In decaffeinated coffees, the caffeine is removed before roasting the coffee. By law decaffeinated coffee cannot contain more than 0.1% of caffeine (Echeverri et al., 2005). These are coffees that have had a resin incorporated either during or after the roasting process, including vanilla, chocolate, strawberry, walnut and amaretto, among others (Farfán Valencia, 2008).

The panel structure and market segmentation previously discussed form the theoretical foundation of the study. They are also used as a guideline to allow a systematic data collection and finally for a consistent analysis, which lead us to accurate results.

## 6 Results

A total of ten different brands of coffee with a high presence in the Colombian market were classified in five market segments. Although the complete market segmentation proposed actually comprises six segments, the category of flavoured and decaffeinated coffees was discarded as this category deals with coffees whose natural characteristics have been modified (ICONTEC, 2011). Two commercial samples were randomly selected from the coffee of each segment studied, and were then evaluated for the nine descriptors presented in Table 1. All the commercial brands were evaluated according to a scale with values from 1 to 10, where 1 represented the 'worst quality' and 10 denoted the 'best quality'. The quality assessment was performed by 32 panellists divided into two groups of undergraduate and graduate students. The experiment consisted of only one replica, meaning the panellists evaluated each product just once. An evaluation form was developed for the project, using as reference the existing technical standards for sensory evaluation of coffee (ICONTEC, 2011), the proposed methodology by CENICAFE (Puerta Quintero, 1997) and the methodology proposed by the SCAA (2009).

The figures are box-plots that show the descriptive statistics for each descriptor. While each commercial brand is represented on the horizontal axis, the grade obtained is given on the vertical axis.

Figures 2 to 10 allow a general comparison of each commercial brand with respect to each descriptor. A total number of nine box-plots are given, relating to the total of nine descriptors defined. The tiny blue square inside each box represents the obtained average. The box-plot analysis is basically an attempt to identify the best and worst evaluated brand, as seen in following lines. Note that for purposes of this analysis, the nine descriptors of the model are assumed having equal weight.

- *Acidity*: the highest variation for this descriptor was found in coffee number 65, and the lowest variation was found in coffee number 72. Number 44 is the most acidic coffee, followed by coffee 35; while number 21 is the least acidic coffee. Note that this descriptor is very important because acidity is closely related with quality. The higher the acidity, the higher the quality (see Figure 2).

- *Bitterness*: Brand number 35 is the most bitter coffee, while the coffee with the highest variation in bitterness is number 87. Coffee 73 on the other hand has the lowest variation. According to the panel, coffee number 72 is the coffee with the least bitterness of the group of evaluated brands (see Figure 3).

- *Aroma*: According to this descriptor coffee number 65 has the strongest aroma, while number 87 is the coffee with the weakest aroma. The coffee with the highest variation in aroma is identified as coffee number 65, while number 77 is the coffee with the lowest variation (see Figure 4).

**Figure 2**  Box-plots for acidity (see online version for colours)

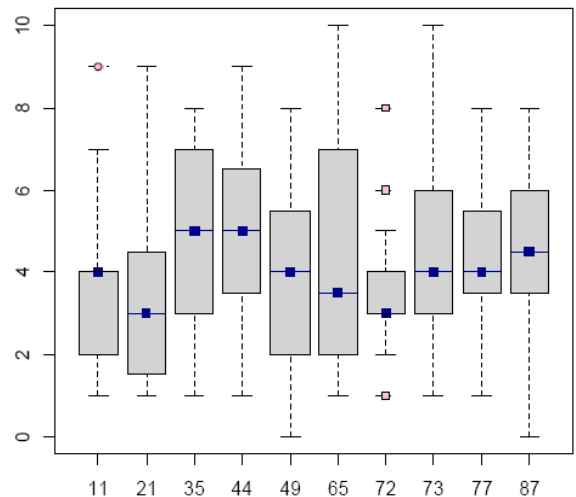

**Figure 3** Box-plots for bitterness (see online version for colours)

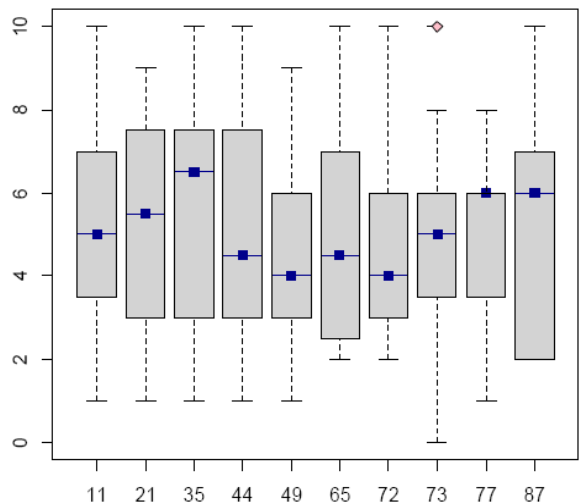

**Figure 4** Box-plots for aroma (see online version for colours)

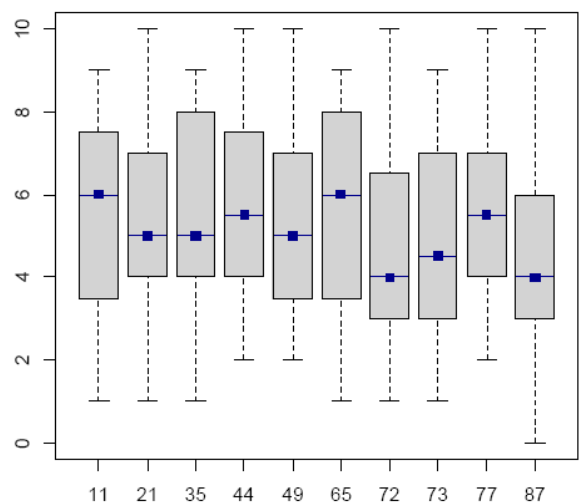

- *Fragrance*: with regard to this descriptor, coffee number 77 was perceived to have the strongest fragrance, while coffee number 87 was the least fragrant. While the coffee with the highest variation was number 72, that with the lowest variation was on coffee 11. This attribute is relevant in terms of consumer preference and perceived quality. In terms of perception, the stronger the fragrance, the higher the quality (see Figure 5).

- *Sweetness*: in general terms this descriptor has a very low rating in coffees numbers 35, 65 and 72, with coffee number 35 the least sweet of this group. Coffee number 49 on the other hand was labelled as the sweetest. In terms of variation, number 49 was also the coffee with the highest variation, while number 11 obtained the lowest variation. It is important to mention that in Colombia coffee is normally consumed after adding sugar. In the case at hand, the coffees were evaluated with no sugar added to the samples, thus explaining the low marks obtained by all brands (see Figure 6).

- *Body*: coffee number 77 was perceived as having the most body, while number 72 was the coffee with the least body. In terms of variation, number 11 obtained the highest variation and coffee 49 the lowest. It is important to reiterate that this attribute is relevant to the overall quality of the product, given that the thickness of the flavour, consistency and thickness of the liquid are important in the preferences of Colombian consumers (see Figure 7).

**Figure 5**   Box-plots for fragrance (see online version for colours)

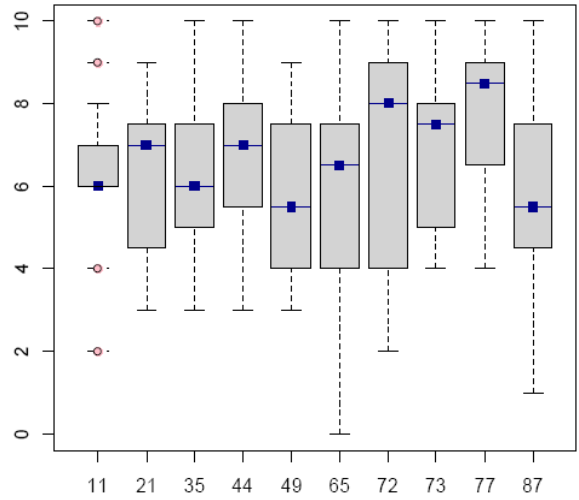

**Figure 6**   Box-plots for sweetness (see online version for colours)

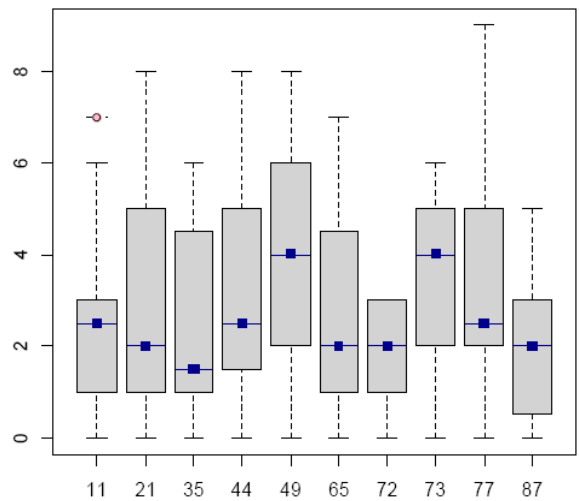

**Figure 7** Box-plots for body (see online version for colours)

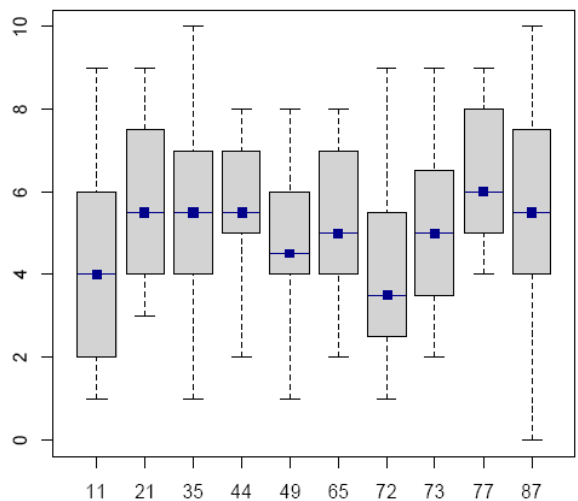

**Figure 8** Box-plots for actron (see online version for colours)

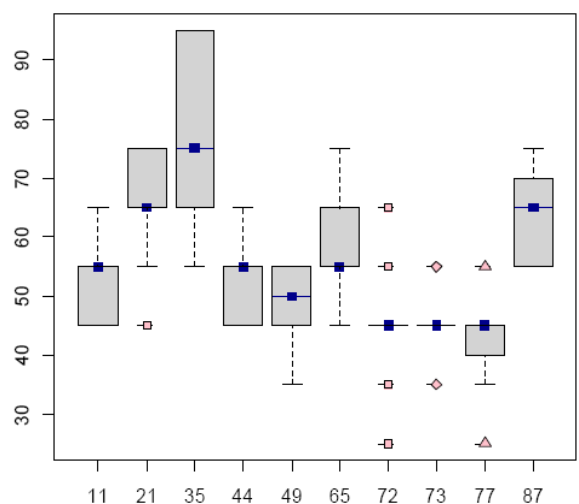

- *Actron*: The lowest variations were identified for coffees 72, 73 and 77, with coffee number 77 the lowest of the entire group. Coffee 35 on the other hand obtained the highest evaluation for this attribute. In terms of variation two coffees obtained the lowest marks for this measure, specifically numbers 72 and 73, meaning that the majority of the judges agree with the respect to this attribute. We therefore find this to be a partial consensus of the group with regard to this attribute, and finally an answer the second question of this research (see Figure 8).

- *Aftertaste*: coffee number 49 achieved the highest average on this descriptor, while coffee number 72 had lowest value. With respect to variation, coffee number 87 had the greatest value while number 72 had the lowest variation. This aspect is important for Colombian customers, because the longer the flavour remains on the mouth, the higher the quality as perceived by consumers.

- *Global impression*: This is the last descriptor evaluated by the judges. The brand with the best global impression was number 44, suggesting that it is the most preferred by the Colombian customers. The brand with the lowest global impression was number 72, suggesting that brand 72 is that least preferred by Colombian consumer.

**Figure 9** Box-plots for aftertaste (see online version for colours)

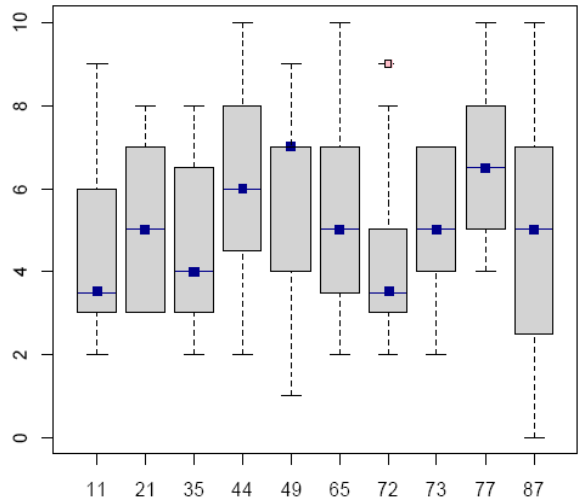

**Figure 10** Box-plots for global impression (see online version for colours)

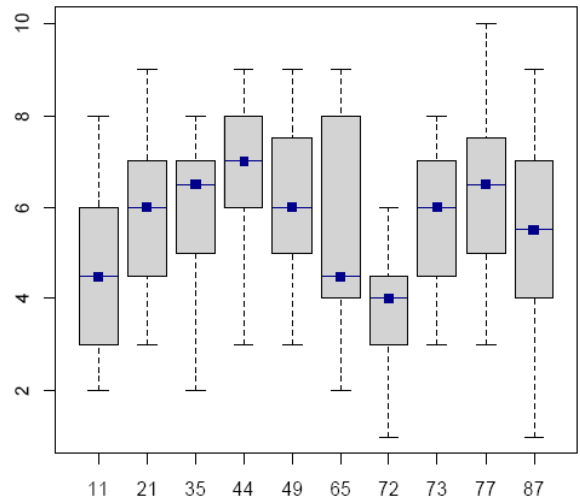

*6.1 Multidimensional analysis*

A multivariate analysis was performed to create a more accurate picture of the Colombian coffee consumer. Using the principal component analysis (PCA) technique, a map was drawn of consumer preferences from the plot loadings. In order to gain a deeper understanding of the preferences of Colombian consumers, grades provided by the judges were analysed by *k-means* cluster using the Euclidean distance (sometimes called a hierarchical cluster analysis method). This method basically relates rows (consumer evaluations) that are closer in the factorial plane with similar consumption preferences. Furthermore, the PCA characterises commercial brands of coffee using a simple tabulation and the Pearson's chi-squared to test similarities among clusters. This is graphically represented on the factorial plane using the first two axes, as presented in Figures 12 and 13. The closer the brands on the factorial plane, the more similar they are. Note that brands 21, 65 and 49 are quite close in Figure 12; therefore, these brands are perceived as similar by the judges. In this form the matrix data obtained from the PCA produces 'loadings' for the coffee brands and 'scores' for consumer evaluations.

According to Figures 11 and 12, coffees located in the upper right quadrant are affected by the overall impression of the aroma, sweetness, fragrance, body and aftertaste. More specifically coffee 44 in the case of the aftertaste. On the other hand, coffee 35 is perceived as the sweetest. Coffees located in the lower right quadrant are affected by bitterness, acidity and agtron, specifically brands 73 and 77. The coffees closest to the horizontal axis are affected more by aftertaste flavour, body and fragrance, while the coffees on the vertical axis are more affected by agtron and sweetness. The coffees labelled with the numbers 35 and 21 are characterised as sweet. Finally, the coffee that obtained the best global impression was coffee 44 (see Figure 13).

**Figure 11** Map of attributes on the factorial plane

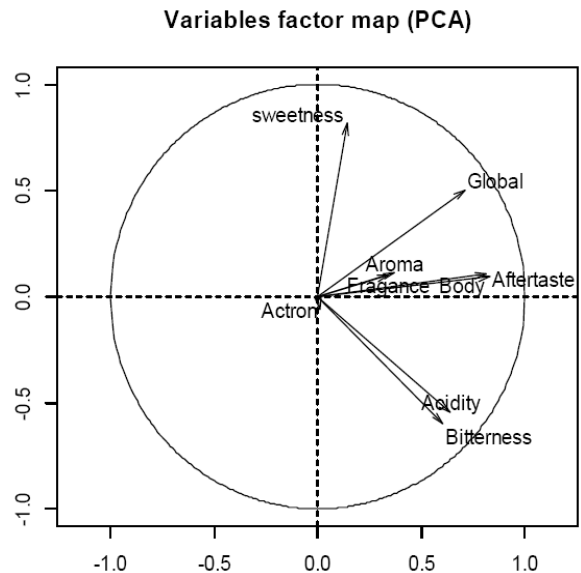

**Figure 12** Map of products on the factorial plane (see online version for colours)

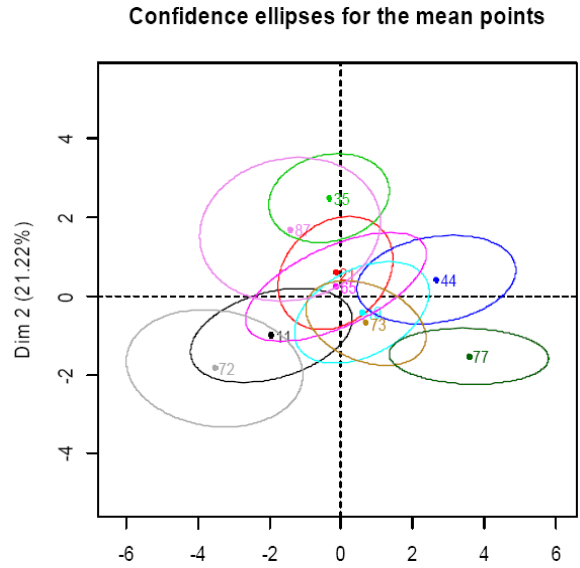

**Figure 13** Map of products on the factorial plane with % of inertia (see online version for colours)

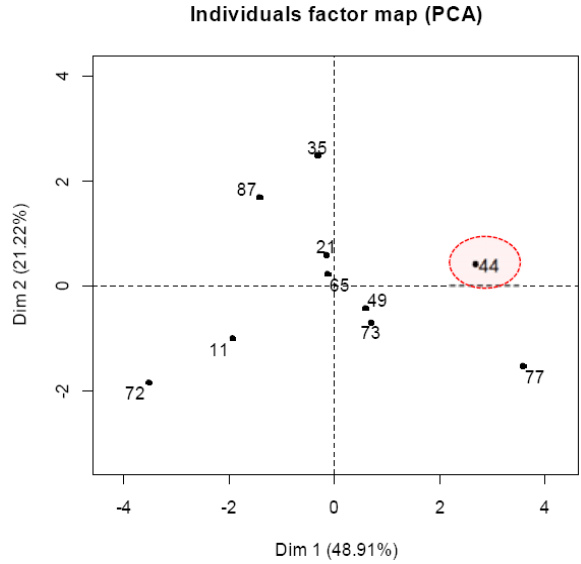

## 7 Conclusions

Consumer preferences, like any other human behaviour, are affected by innumerable interrelated factors. In the particular case of the hot beverages industry, attributes that consumers perceive through their senses are of strategic importance. It is evident that companies that expect to create competitive advantages should make a deeper investigation into sensory attributes, in order to align their products with the consumer preferences.

This paper discusses nine sensory attributes of hot coffee, attributes that based on our review of literature are relevant to the Colombian consumer. A group of ten brands of coffee that predominate in the Colombian market were investigated under the given attributes. A panel composed of 32 individuals made a single evaluation of each coffee. The name of each commercial brand is not published here, as that is considered to be sensitive information. However, we are willing to provide the full results on request sent to the main author's e-mail address. According to our results, there is a tendency in Colombia to consume coffees that have a strong fragrance, aroma, body and aftertaste. Colombian consumers similarly tend to prefer sweeter coffees, rather than bitter. These features, as discussed earlier, should be observed by commercial coffee brands if they expect to be successful on Colombian market. For instance, the product identified as number 44 received the highest overall impression, while product 72 was identified as the worst in terms of the mentioned attributes. As seen in Figure 11, attributes such as aroma, fragrance, body and aftertaste are strongly related with the global impression.

At this point, it is important to highlight the limitations of these conclusions. Since the study included 32 panellists, each of whom evaluated a coffee brand only once, results might be altered by variables such as weather conditions, education level and age, among others. Further research should include socio-demographic variables and replicas during different months of the year.

Sensory evaluation is a discipline that is growing in importance due its practical and convenient results for the food and beverage industries. This project is a limited, but illustrative example of the suitability of this methodology to align features of the beverages to the voice of the customers.

## References

Breslow, N.E. (1996) *Generalized Linear Models: Checking Assumptions and Strengthening Conclusions*, 31 July, Department of Biostatistisc University of Washington [online] http://citeseerx.ist.psu.edu/viewdoc/download?doi=10.1.1.50.6105&rep=rep1&type=pdf (accessed 21 January 2013).

Calcagno, V. and Mazancourt, C. (2010) *glmulti: An R Package for Easy Automated Model Selection with (Generalized) Linear Models*, May [online] http://www.jstatsoft.org/ (accessed 6 February 2014).

Calí, M. (2002) *Análisis sensorial de los alimentos. Fruticultura & Diversificación INTA*, EEA Alto Valle [online] http://www.inta.gov.ar/altovalle/info/biblo/rompecabezas/pdfs/fyd48_entrev.pdf (accessed 10 March 2014).

Calvo Fernández, S. (1997) *Factores determinantes de la calidad percibida: Influencia en la decisión de compra* [online] http://eprints.ucm.es/3717/.UCM Ed. Madrid (accessed 21 March 2014).

Césari, M., Césari, R., Gámbaro, A. and Arnau, E. (2013) 'Análisis de datos provenientes de pruebas sensoriales del vino, utilizando la lógica borrosa', paper presented at the *Séptimo Encuentro de Investigadores y Docentes de Ingeniería*, Los Reyunos, Mendoza, Argentina [online] http://www.researchgate.net/profile/Matilde_Ines_Cesari/publication/255880186_Acesari1/links/0c960520cc5d69451c000000 (accessed 22 February 2014).

Couillerot, C. (2000) *The Protected Designations of Origin*, Institute of Rural economy of ETH from Zurich [online] http://www.aoc.igp.com/Aopgb/haopgb.htm (accessed 13 April 2014).

Davids, K. (2010) *Saying Coffee*, November, The Naming Revolution [online] http://www.roastmagazine.com/resources/Articles/Roast_NovDec10_SayingCoffee.pdf (accessed 21 April 2014).

Echeverri, D., Buitrago, L., Montes, F., Mejía, I. and González, M.d. (2005) 'Café para cardiólogos', *Revista Colombiana de Cardiología*, Vol. 11, No. 8, pp.357–365.

Farfán Valencia, F. (2008) *CENICAFE*, 22 September [online] http://www.cenicafe.org/es/documents/LibroSistemasProduccionCapitulo10.pdf (accessed 7 May 2014).

FEDECAFE (2006) *ASPECTOS DE CALIDAD DEL CAFÉ PARA LA INDUSTRIA TORREFACTORA NACIONAL*, 24 August [online] http://www.iue.edu.co/portal/images/negocios_internacionales/cafe/LACALIDADENLAINDUSTRIADELCAFE.pdf (accessed 6 March 2013).

Food Industry (2014) *proceso de elaboracion del café*, 15 April, de cafe tostado [online] http://ben.upc.es/documents/eso/aliments/html/estimulantes-4.html (accessed 22 February 2014).

Garvin, D.A. (1984) 'What does 'product quality' really means?', *Sloan Management Review*, Vol. 26, No. 1, pp.25–43 [online]http://www.oqrm.org/English/What_does_product_quality_really_means.pdf (accessed 21 March 2013).

Granados, L. and Alvarez, C. (2002) 'Variabilidad de establecer el sistema de denominacion de origen de los productos agroalimentarios en Costa Rica', *Aostarricense*, Vol. 26, No. 1, pp.63–72 [online] http://www.redalyc.org/pdf/436/43626106.pdf (accessed 21 April 2014).

ICONTEC (2011) 'Norma Técnica Colombiana NTC 4883', *Analisis Sensorial. Café. Metodología para analisis sensorial cuantitativo descriptivo del café*, ICONTEC, Bogotá, Colombia.

INFOCAFES (2009) *Caracteristicas sensoriales*, October [online] http://www.infocafes.com/descargas/biblioteca/24.pdf (accessed 22 February 2014).

INFUSIONISTS (2014) *Cata de café* [online] http://infusionistas.com/cata-de-cafe/ (accessed 6 March 2013).

Lacaze, V., Rodriguez, E.M. and Lupín, B. (2007) 'Alimentos diferenciados', *FACES*, Vol. 13, No. 28, pp.7–34.

Nelder, J.A. and Wedderburn, R.M. (1972) 'Generalized linear models', *Journal of the Royal Statistical Society. Series A (General)*, Vol. 135, No. 3, p.370–384.

Portfolio (2012) *Crece el consumo de café en Colombia*, 13 March [online] http://www.portafolio.co/economia/crece-el-consumo-cafe-colombia (accessed 21 April 2014).

Puerta Quintero, G.I. (1988) *Calidad en la taza de las variedades de Coffea arabica L. cultivadas en Colombia*, 1 December, CENICAFE [online] http://biblioteca.cenicafe.org/handle/10778/64 (accessed 10 March 2014).

Puerta Quintero, G.I. (1997) *Escala para la Evaluacion de la calidad de la bebida de café verde Coffea arabica, procesado por via húmeda*, 1 January, CENICAFE [online] http://biblioteca.cenicafe.org/handle/10778/62 (accessed 6 March 2013).

Puerta Quintero, G.I. (1998) *Calidad En Taza De Las Variedades De Coffea arabica L*, CENICAFE, pp.265–278, Obtenido de cultivadas en colombia.

Puerta Quintero, G.I. (2009) 'Los catadores de café', *Avances técnicos 381 Cenicafé*, pp.1–12.

Revista pyme.com (2013) *En #MiDíaDelCafé, un análisis del consumo de café en Colombia*, 27 June [online] http://www.revistapym.com.co/destacados/verdad-acerca-cafe-colombia-ofertademanda-publicidad-consumo (accessed 21 April 2014).

Specialty Coffee Association of America (SCAA) (2009) *green-coffee-color*, 21 November [online] http://www.scaa.org/PDF/resources/green-coffee-color.pdf (accessed 15 March 2014).

SensoMiner (2015) *SENSOMINER*, 20 January [online] http://sensominer.free.fr/ (accessed 2 January 2015).

Steenkamp, J-B.E. (1989) *Product Quality: An Investigation into the Concept and How It Is Perceived by Consumers: Wageningen*, Van Gorcum Assen Press, Netherlands.

The Coffee Guide (2014) *Medicion del color del café tostado*, Centro de Comercio Internacional [online] http://www.laguiadelcafe.org/guia-del-cafe/calidad-del-cafe/Medicion-del-color-del-cafetostado/?menuID=3377 (accessed 5 May 2014).

USAID (2005) *Normas y estandares de catación*, 29 August, Para la region de centroamérica [online] http://pdf.usaid.gov/pdf_docs/pnadg946.pdf (accessed 15 April 2014).

Vaclavik, V.A. (1998) *Fundamentos de ciencias de los alimentos*, Editorial Acribia S:A, Zaragoza.

Zapata Gómez, A. and Sarache Castro, W.A. (2014) *Mejoramiento de la calidad del café soluble utilizando el método Taguchi*, Obtenido de Ingeniare. Revista chilena de ingeniería, January, http://www.scielo.cl/scielo.php?script=sci_arttext&pid=S0718-33052014000100011&lng=es&nrm=iso (accessed 16 March 2014).